\documentclass[aps,twocolumn,superscriptaddress,groupedaddress,nofootinbib]{revtex4-1}
\usepackage{graphicx}
\usepackage{dcolumn}   
\usepackage{bm}       
\usepackage{amssymb}
\usepackage{amsmath}
\usepackage[toc,page]{appendix}
\usepackage{braket}
\newcommand{\del}{\partial}

\begin{document}

\title{ Geometry of Two-Sheeted Spacetime Solutions }

\author{Sandipan Sengupta}
\email{sandipan422@gmail.com\\sandipan@phy.iitkgp.ac.in}
\affiliation{Department of Physics and Centre for Theoretical Studies, Indian Institute of Technology Kharagpur, Kharagpur-721302, INDIA}

\begin{abstract}

In contrast to Einstein's theory, the first order formulation of gravity turns out to be a natural habitat for double-sheeted spacetime solutions which satisfy the vacuum field equations everywhere. These bridge-like geometries exhibit degenerate tetrads at their core that separates the two sheets. Here we study the geodesics of these solutions and elucidate their causal structure. These spacetimes emerge as a classical realization of a two-universe solution in pure gravity.  We also find the angle of deflection of light propagating in a bridge geometry. From this, we conclude that this spacetime would be indistinguishable from the Schwarzschild exterior when observed from asymptotia.


\end{abstract}

\maketitle

\section{Introduction}
Double sheeted spacetimes made their first serious appearance within the realm of general relativity through the work of Einstein and Rosen \cite{rosen}, in what could be seen as a remarkable yet failed attempt to build a geometric model of elementary particles. From a modern perspective, the Einstein-Rosen bridge is essentially a non-traversable wormhole. Even though this configuration may be represented as a part of the maximally extended Schwarzschild geometry in Kruskal-Szekeres coordinates \cite{kruskal,szekeres}, they are not solutions of the Einstein's equations in vacuum. The geometric and topological properties of these configurations have been the centre of intense investigation over the years \cite{fuller,rindler}. Here, we focus on two-sheeted spacetimes of a different kind, which have no analogue in the Einsteinian gravity. These are explicit solutions to the first order equations of motion in vacuum \cite{tseytlin} and have been constructed only recently \cite{sengupta}.  These are associated with a metric whose determinant varies continuously between zero and non-zero values over different regions of the same spacetime. These represent what have been named as spacetime-bridge geometries, where the noninvertible tetrad phase ($\det e_\mu^I=0$) defines the core and the invertible phase ($\det e_\mu^I\neq 0$) defines the two spacetime sheets away from it. 
The bridge spacetimes are regular in the sense that the basic gauge-covariant fields (tetrad, torsion and field-strength) are all continuous across the boundaries between the degenerate and nondegenerate phases and the field-strength components are finite everywhere. 

Here our goal is to understand the geometric properties of the spacetime-bridge geometries. To be precise, we find the geodesics of these spacetimes, and elucidate their causal and completeness properties. Such a study is essential to understand the relevance of this set of non-Einsteinian vacuum solutions in the context of classical gravity theory.

In the next section, we present a set of spacetime bridge solutions, which are constructed to be free of torsion. This should not be seen as a generic feature though, since solutions of the first order equations of motion may possess torsion in general \cite{tseytlin,kaul,kaul1,sengupta}. 
The geodesic equations in the regions associated with invertible as well as noninvertible tetrads are set up next and their solutions are discussed.  We then go on to find the angle of deflection of light propagating in a bridge spacetime and comment on whether it is possible to distinguish this geometry from the (exterior) Schwarzschild spacetime, which is the unique spherically symmetric solution in the Einsteinian phase. The final section contains a summary of the main results along with a few remarks on the future prospects. 

\section{Double-sheeted solutions}
The first order equations of motion, which are obtained from the variation of the Hilbert-Palatini Lagrangian density, are given by \cite{tseytlin}:
\begin{eqnarray}\label{eom1}
e_{[\mu}^{[K} D_{\nu}(\omega)e_{\alpha]}^{L]}=0,~e_{[\nu}^{[J} R_{\alpha\beta]}^{~KL]}(\omega)=0~
\end{eqnarray}
The two sets of equations above correspond to the independent variations with respect to the two independent fields, namely, the tetrad and spin-connection.
Our analysis here would be based on a set of two-sheeted bridge-spacetimes, to be constructed as explicit solutions to these field equations and having zero torsion:
\begin{eqnarray}
 D_{[\mu}(\omega)e_{\nu]}^I=\del_{[\mu} e_{\nu]}^I+\omega_{[\mu}^{~IL}e_{\nu]L}=0.
\end{eqnarray}

Let us choose the coordinates ($t,v,\theta,\phi$), where the full spacetime is described by the ranges $t\in(-\infty,\infty),~v\in(-\infty,\infty),~\theta\in[0,\pi],~\phi\in[0,2\pi]$. Each $t=const.,~v=const.$ slice represents a spatial two-sphere.  The two spacetime sheets are represented by the patches  $v>\epsilon$ and $v<-\epsilon$ respectively, assumed to carry (static) metrics of the general form\cite{sengupta}:
\begin{eqnarray}\label{sc}
 ds^2
 =&-&\left[\frac{f^2(v)}{f^2(v)+2M}\right]dt^2 + 4\left[f^2(v)+2M\right]f^{'2}(v) dv^2
 \nonumber\\
 &+& \left[f^2(v)+2M\right]^2 d\Omega^2~ \equiv ~g_{\mu\nu} d x^\mu d x^\nu
\end{eqnarray}
The function $f(v)$ is assumed to be monotonic (at $|v|>\epsilon$) and it satisfies:
\begin{eqnarray}\label{f(u)}
f(\pm\epsilon)=0=f'(\pm\epsilon),~f(|v|)\rightarrow \infty \mathrm{~as~}|v|\rightarrow \infty
\end{eqnarray}
At the intermediate (`bridge') region $|v|\leq \epsilon$, the metric is degenerate:
\begin{eqnarray}\label{G}
ds^2
&=& 0+\sigma F^2(v) dv^2 + H^2(v)\left[d\theta^2+\mathrm{sin}^2 \theta d\phi^2\right]\nonumber\\
&\equiv& \hat{g}_{\mu\nu} d x^\mu d x^\nu,
\end{eqnarray}
where $F(v)$ and $H(v)$ are two arbitrary functions to be solved using the equations of motion and $\sigma=\pm 1$. The internal metric $\eta_{IJ}=diag[-1,1,1,1]$ is defined to be the same everywhere.
One may note that in the regions $|v|>\epsilon$, the metric above can be brought to the 
Einstein-Rosen form \cite{rosen} upon a change of coordinates: $f(v)=u$. However, such an equivalence fails at the phase boundaries $v=\pm \epsilon$ where the reparametrization becomes singular, as well as within the core at $|v|<\epsilon$. Singular coordinate transformations similar to the above have also been considered earlier in constructing solutions to the canonical constraints of the Ashtekar formulation, first by Bengtsson \cite{bengtsson} and subsequently by others \cite{bengtsson1,madhavan}. 

The general method to construct the bridge solutions based on the above metrics has been discussed in ref.\cite{sengupta}. Such vacuum solutions  carry nontrivial torsion at their core. The torsion-free solutions to be presented here are different from those in general.

\subsection*{Field configuration at the invertible phase}

Given the four-metric (\ref{sc}), we can obtain the (torsionless) spin-connection fields $\omega_\mu^{~IJ}(e)=\frac{1}{2}[e^{\rho I}\del^{}_{[\mu}e_{\rho]}^J
-e^{\rho J}\del^{}_{[\mu}e_{\rho]}^I -  e_\mu^L e^{\rho I} e^{\sigma J}
\del^{}_{[\rho}e_{\sigma]L}]$, whose nonvanishing components are given by:
\begin{eqnarray}\label{omega}
\omega_t^{~01}&=&\frac{M}{[f^2(v)+2M]^2},~\omega_\theta^{~12}=-\frac{f(v)}{[f^2(v)+2M]^{\frac{1}{2}}},\nonumber\\ \omega_\phi^{~23}&=&-\cos\theta,~
\omega_\phi^{~31}=\frac{f(v)}{[f^2(v)+2M]^{\frac{1}{2}}} \sin\theta~.
\end{eqnarray}
These lead to the following nontrivial components for the curvature two-form $R^{~IJ}_{\mu\nu}(\omega)=\del_{[\mu}\omega_{\nu]}^{~IJ}+\omega_{[\mu}^{~IL}\omega_{\nu]L}^{~~~J}$:
\begin{eqnarray}\label{RR}
R_{tv}^{~01}&=&\frac{4M f(v)f'(v)}{[f^2(v)+2M]^{3}},R_{t\theta}^{~02}=-\frac{M f(v)}{[f^2(v)+2M]^{\frac{5}{2}}},\nonumber\\R_{t\phi}^{~03}&=&-\frac{M f(v)}{[f^2(v)+2M]^{\frac{5}{2}}}\sin\theta,
R_{v\theta}^{~12}=-\frac{2M f'(v)}{[f^2(v)+2M]^{\frac{3}{2}}},\nonumber\\R_{\theta\phi}^{~23}&=&
\frac{2M }{[f^2(v)+2M]}\sin\theta ,R_{\phi v}^{~31}=-\frac{2M f'(v)}{[f^2(v)+2M]^{\frac{3}{2}}}\sin\theta \nonumber\\
~~
\end{eqnarray}
This configuration defining the regions at $|v|>\epsilon$ satisfies the first order equations of motion, which are equivalent to the Einstein's equations in vacuum (as the tetrad is invertible).

An equivalent description of the above field configuration may also be provided in terms of a set of variables (i.e., metric, affine-connection and field-strength) that are insensitive to the internal $SO(3,1)$ gauge rotations. These are defined as :
\begin{eqnarray}\label{gamma1}
\Gamma_{\alpha\beta\rho}&=&\Gamma_{\alpha\beta}^{~~\sigma}g_{\rho\sigma}=\frac{1}{2}\left[\del_{\alpha} g_{\beta\rho}+  \del_{\beta} g_{\alpha\rho}-  \del_{\rho}g_{\alpha\beta}\right]\nonumber\\
R_{\alpha \beta\rho\sigma}&=&R_{\alpha\beta}^{~~IJ}e_{\rho I} e_{\sigma J}
\end{eqnarray}
In view of their relevance for a later part of this analysis, let us list the nontrivial affine connection components associated with the invertible phase:
\begin{eqnarray}\label{gamma2}
&&\Gamma_{ttv}=\frac{2Mf(v)f'(v)}{\left[f^2(v)+2M\right]^2},~\Gamma_{tvt}=-\frac{2Mf(v)f'(v)}{\left[f^2(v)+2M\right]^2}=\Gamma_{vtt},\nonumber\\&&\Gamma_{vvv}=2\del_v \left(\left[f^2(v)+2M\right]f^{'2}(v)\right),\nonumber\\
&&\Gamma_{\theta\theta v}=-2f(v)f'(v)\left[f^2(v)+2M\right],\nonumber\\&&\Gamma_{\phi\phi v}=-2f(v)f'(v)\left[f^2(v)+2M\right] \sin^2 \theta,\nonumber\\
&&\Gamma_{v\theta\theta}=2f(v)f'(v)\left[f^2(v)+2M\right]=\Gamma_{\theta v \theta},\nonumber\\
&&\Gamma_{v\phi\phi}=2f(v)f'(v)\left[f^2(v)+2M\right]\sin^2 \theta=\Gamma_{\phi v\phi},\nonumber\\
&&\Gamma_{\phi\phi\theta}=-\left[f^2(v)+2M\right]^2 \sin\theta\cos\theta,\nonumber\\
&&\Gamma_{\theta\phi\phi}=\left[f^2(v)+2M\right]^2 \sin\theta \cos\theta=\Gamma_{\phi\theta\phi}~.
\end{eqnarray}

 \subsection*{Field configuration at the noninvertible phase}
 For the region $|v|\leq\epsilon$, the nontrivial triad fields may be read off from the metric $\hat{g}_{\mu\nu}$ in  (\ref{G}) to be:
\begin{eqnarray}
\hat{e}^0=0,~  \hat{e}^1=\sqrt{\sigma} F(v) dv,~  \hat{e}^2=H(v)d\theta,~   \hat{e}^3=H(v)\sin\theta d\phi
\end{eqnarray} 
With these, we look for a solution with the following ansatz for the spin-connection fields:
\begin{eqnarray}\label{omega*}
&&\hat{\omega}_t^{~0i}=0=\hat{\omega}_t^{~ij},~\hat{\omega}_a^{~0i}=0,\nonumber\\
&&\hat{\omega}_a^{~ij}=\frac{1}{2}\left[\hat{e}^{bi}\del^{}_{[a}\hat{e}_{b]}^j
-\hat{e}^{bj}\del^{}_{[a}\hat{e}_{b]}^i -  \hat{e}_a^l \hat{e}^{bi} \hat{e}^{cj}
\del^{}_{[b}\hat{e}_{c]l}\right]
\end{eqnarray}
where $\hat{e}_a^i$ are the triads associated with the nondegenerate three-subspace:
\begin{eqnarray}\label{e}
\hat{e}^i_a=\left(\begin{array}{ccc}
F(v) & 0 & 0\\
0 & H(v) & 0\\
0 & 0 & H(v)\sin\theta  \end{array}\right) 
\end{eqnarray} 
Using these, we define $\hat{h}_{ab}=\hat{e}_a^i \hat{e}_{bi}$ to be the three-metric induced at the $(v,\theta,\phi)$ submanifold with:
\begin{eqnarray*}
\hat{h}^{ab}\hat{h}_{bc}=\delta^a_c,~\hat{h}^{ab}\hat{h}_{ab}=3.
\end{eqnarray*} 
The spin-connection (\ref{omega*}) have zero torsion by construction and may be rewritten as:
\begin{eqnarray}\label{omegafull}
&&\hat{\omega}^{0i}=0,~
\hat{\omega}^{12}=-\frac{H'(v)}{\sqrt{\sigma}F(v)} d\theta,~
\hat{\omega}^{23}=-\cos\theta d\phi,\nonumber\\
&&\hat{\omega}^{31}=\frac{H'(v)}{\sqrt{\sigma}F(v)} \sin\theta d\phi~.
\end{eqnarray}
The nonvanishing components of the affine connection $\hat{\Gamma}_{\alpha\beta\rho}=\hat{\Gamma}_{\alpha\beta}^{~~\sigma} \hat{g}_{\rho\sigma}=\frac{1}{2}\left[\del_{\alpha} \hat{g}_{\beta\rho}+  \del_{\beta} \hat{g}_{\alpha\rho}-  \del_{\rho}\hat{g}_{\alpha\beta}\right]$
 are also listed below:
\begin{eqnarray}\label{gamma4}
&&\hat{\Gamma}_{vvv}=\frac{\sigma}{2}\del_v F^2 (v),\hat{\Gamma}_{\theta\theta v}=-\frac{1}{2}\del_v H^2 (v)+\sigma \mu(v)F(v)H^2(v),\nonumber\\
&&\hat{\Gamma}_{v\theta\theta}=\frac{1}{2}\del_v H^2(v)=\hat{\Gamma}_{\theta v \theta},\nonumber\\&&\hat{\Gamma}_{\phi\phi v}=-\left[\frac{1}{2}\del_v H^2(v)+\sigma \mu(v)F(v)H^2(v) \right]\sin^2 \theta,\nonumber\\
&&\hat{\Gamma}_{\phi\phi\theta}=-H^2 (v) \sin\theta\cos\theta=-\hat{\Gamma}_{\theta\phi\phi}=-\hat{\Gamma}_{\phi\theta\phi}
\end{eqnarray}
The field-strength components corresponding to the spin-connection fields (\ref{omegafull}) are evaluated to be:
\begin{eqnarray}\label{R-simple}
\hat{R}^{0i} (\hat{\omega})&=& 0,~~\hat{R}^{12}(\hat{\omega})=-\frac{1}{\sqrt{\sigma}}\left[\frac{H'(v)}{F(v)}\right]' dv\wedge d\theta ,\nonumber\\
\hat{R}^{23}(\hat{\omega})&=&\left[1-\sigma \frac{H^{'2}(v)}{F^2(v)}\right]\sin\theta ~d\theta\wedge d\phi ,\nonumber\\
\hat{R}^{31}(\hat{\omega})&=&-\frac{1}{\sqrt{\sigma}}\left[\frac{H'(v)}{F(v)}\right]'\sin\theta ~d\phi\wedge dv
\end{eqnarray}
For $\sigma=-1$, some of the fields above are imaginary. However, their $SO(3,1)$-invariant counterparts are all manifestly real.

Since the degenerate configuration here is associated with vanishing torsion, it trivially solves the first among the set of equations of motion (\ref{eom1}). The remaining one is also satisfied provided the metric functions ($F(v),H(v)$) are constrained as:
\begin{eqnarray}\label{master}
\hat{e}_{[a}^{[i} \hat{R}_{bc]}^{~~jk]}=0=\frac{H'^2(v)}{F(v)}+2 \left[\frac{H'(v)}{F(v)}\right]'H(v)-\sigma F(v)
\end{eqnarray}
This can be solved for the metric functions in terms of a (real) function $\Omega(v)=\sqrt{2M}\left(\frac{H'(v)}{F(v)}\right)$ as:
\begin{eqnarray}\label{FH}
F(v)=-\frac{\sigma\sqrt{8M} \Omega'(v)}{\left[\frac{\Omega^2(v)}{2M}-\sigma\right]^2},~H(v)=\frac{2M}{\left[\frac{\Omega^2(v)}{2M}-\sigma\right]}
\end{eqnarray} 
Here we have used the continuity of the metric across the boundaries $v=\pm \epsilon$ to fix the integration constants. The expressions of the exterior and interior field-strength components in eq.(\ref{RR}) and (\ref{R-simple}) reveal that these fields (and in general, all the gauge-invariant fields) are also continuous provided:
\begin{eqnarray*}
\Omega(\pm\epsilon)=0=\Omega'(\pm\epsilon)~.
\end{eqnarray*}
With this, the metric (\ref{G}) within the bridge may be rewritten as:
\begin{eqnarray}\label{f-metric}
ds^2=0+\frac{8\sigma M \Omega^{'2}(v)}{\left[\frac{\Omega^2(v)}{2M}-\sigma\right]^4 } dv^2
+\frac{4M^2}{\left[\frac{\Omega^2(v)}{2M}-\sigma\right]^2}(d\theta^2+\sin^2\theta d\phi^2)\nonumber\\
~~
\end{eqnarray}

Evidently, the field configuration within the degenerate core at $|v|< \epsilon$ is completely insensitive to the null coordinate $t$. 
This fact allows us to provide a three-geometric interpretation to this region, even though it is originally four-dimensional. Such a `three-geometry' is manifestly regular. This is so because the associated three-curvature scalars, constructed out of $\bar{\omega}_a^{~ij}(\hat{e})$ determined purely by the triads $\hat{e}_a^i$, are nonsingular. For instance, the linear and quadratic scalars are found to be:
\begin{eqnarray}\label{Rbar}
\bar{R}=0,~~ \hat{h}^{ac} \hat{h}^{bd} \bar{R}_{ab}^{~~ij} 
\bar{R}_{cdij}=\frac{3}{8 M^4} \left[\frac{\Omega^2(v)}{2M}-\sigma\right]^6
\end{eqnarray}

\section{ Geodesics}
To begin with, we first analyze the geodesic equations for $\sigma=-1$. In this case, the coordinate $v$, which is spacelike away from the bridge, becomes timelike within it.

\subsection*{ Region away from the bridge:}
The geodesic equation is defined as:
\begin{eqnarray}
u^\alpha {\cal{D}}_\alpha u^\beta:=u^\alpha\del_\alpha u^\beta+\Gamma^{~~\beta}_{\alpha\rho} u^\alpha u^\rho=0
\end{eqnarray}
where $u^\alpha=\frac{dx^\alpha}{d\tau}$ is the tangent vector along an affinely parametrized curve $x^\alpha(\tau)$ and $\Gamma^{~~\rho}_{\alpha\beta}$ are given by eq.(\ref{gamma1}). The symmetry of the metric (\ref{G}) suggests that we may as well project the geodesic equations at the equatorial plane $\theta=\frac{\pi}{2}$ and look for the solutions.

The equations of motion for a test particle with dynamical coordinates ($t,v,\phi$) are found to be, respectively:
\begin{eqnarray}\label{geod}
&&\frac{f^2}{f^2+2M} \dot{t}=E,\nonumber\\
&& 4(f^2+2M) f^{'2}\dot{v}^2-\frac{f^2}{f^2+2M} \dot{t}^2+(f^2+2M)^2\dot{\phi}^2 +k=0,\nonumber\\
&&(f^2+2M)^2\dot{\phi}=L
\end{eqnarray}
where, $E$ and $L$ are the constants of motion resulting from the $t$ and $\phi$-independence of the metric coefficients. The values $k=1,0,-1$ define the timelike, null and spacelike geodesics, respectively. Here we shall be concerned with the timelike and null geodesics only, since these encode the physical trajectories of massive and massless particles.
Eliminating $\dot{t}$ and $\dot{\phi}$, we can rewrite the three equations above as:
\begin{eqnarray*}\label{veq}
 \dot{v}^2=\frac{1}{4f^{'2}}\left[\frac{E^2}{f^2}-\frac{L^2}{(f^2+2M)^3} -\frac{k}{(f^2+2M)}\right],
\end{eqnarray*}
which can be solved for the affine parameter as:
\begin{eqnarray}\label{veq}
 \lambda=\pm 2\int dv \frac{ f'}{\left[\frac{E^2}{f^2}-\frac{L^2}{(f^2+2M)^3} -\frac{k}{(f^2+2M)}\right]^{\frac{1}{2}}}
\end{eqnarray}
In the above, the signs $\pm$ refer to the outgoing and ingoing curves, respectively.
\vspace{.2cm}

For the sake of simplicity, let us consider the geodesics with $L=0$. The solution for the null  and timelike curves read, respectively:
\begin{eqnarray}\label{geo1}
\lambda &=&\pm \frac{f^2(v)}{E}+\mathrm{const.} ~~(k=0),\nonumber\\
\lambda &=& \pm \frac{2}{3\sqrt{2M}}(f^2+2M)^{\frac{3}{2}}+ \mathrm{const.}~~  (k=1) 
\end{eqnarray}
In the last line above, we have assumed the constant (energy per unit mass) $E$ to be unity, implying that  the proper time ($\lambda$) is equivalent the coordinate time ($t$) at asymptotia ($v\rightarrow \pm \infty$). 

\subsection*{Region within the bridge:}
The geodesic equations at the bridge region $|v|< \epsilon$ are
defined as:
\begin{eqnarray}
u^\alpha \hat{{\cal{D}}}_\alpha u^\beta:=u^\alpha\del_\alpha u^\beta+\hat{\Gamma}^{~~\beta}_{\alpha\rho} u^\alpha u^\rho=0
\end{eqnarray}
where $\hat{\Gamma}^{~~\rho}_{\alpha\beta}$ has been defined in terms of the metric $\hat{g}_{\mu\nu}$ and the affine connection $\hat{\Gamma}_{\alpha\beta\rho}$ in the previous section. Due to the noninvertibility of the metric, not all the components of $\hat{\Gamma}^{~~\rho}_{\alpha\beta}$ are determined uniquely. The constraint $\hat{g}_{t\mu}=0$ implies the following solution for the affine connection fields:
\begin{eqnarray}\label{gamma-soln}
\hat{\Gamma}^{~~a}_{\alpha\beta}=~\hat{h}^{ab}\hat{\Gamma}_{\alpha\beta b},~~\hat{\Gamma}^{~~t}_{\alpha\beta}=\mathrm{Arbitrary}
\end{eqnarray}
The indeterminacy of the $\hat{\Gamma}^{~~t}_{\alpha\beta}$ is a reflection of the fact that the motion along the null coordinate $t$ within the bridge has no physical content. 

At the equatorial plane, the geodesic equations for the dynamical coordinates $(v,\phi)$ read:
\begin{eqnarray}\label{geod1}
&&\ddot{v}+\hat{h}^{vv}\hat{\Gamma}_{\alpha\beta v} \dot{x}^\alpha \dot{x}^\beta\Rightarrow ~~~\ddot{v}+\frac{\sigma}{F^2}[\sigma FF' \dot{v}^2-HH'\dot{\phi}^2]=0,\nonumber\\
&& \ddot{\phi}+\hat{h}^{\phi\phi}\hat{\Gamma}_{\alpha\beta \phi} \dot{x}^\alpha \dot{x}^\beta\Rightarrow ~~\ddot{\phi}+2\left(\frac{H'}{H}\right) \dot{v}\dot{\phi}=0
\end{eqnarray}
These equations may be rewritten as:
\begin{eqnarray*}
\dot{v}^2=\frac{1}{F^2}\left(\frac{L^2}{H^2}+k\right)~,
\end{eqnarray*}
with the solution:
\begin{eqnarray}\label{tau}
\lambda=\pm \int dv \frac{F}{\left(\frac{L^2}{H^2}+k\right)^{\frac{1}{2}}}.
\end{eqnarray}
Here $L$ is the same constant of motion as introduced in eq.(\ref{geod}).
The solution for $L=0$ geodesics are given by:
\begin{eqnarray}\label{geo1}
v&=&\mathrm{const.} ~~(k=0),\nonumber\\
\lambda &=&\pm \int dv~ F
 =\pm 2M \left[\frac{\frac{\Omega}{\sqrt{2M}}}{\left(\frac{\Omega^2}{2M}+1\right)}+\arctan\left(\frac{\Omega}{\sqrt{2M}}\right) \right]\nonumber\\~&&+\mathrm{const.}~~(k=1).
\end{eqnarray}

In the light of the solutions for the geodesics obtained above, let us now outline a few important properties of the spacetime-bridge geometries. Along any timelike geodesic, the components of the tangent vector diverge as one approaches the phase boundaries $v=\pm \epsilon$ from either $v<-\epsilon$ or $v>\epsilon$. This implies that these geodesics cannot be continued across this hypersurface. The two spacetime sheets at $v>\epsilon$ and $v<-\epsilon$ are thus causally disconnected and may be interpreted as two `universes' separated by an acausal `bridge' made up of zero determinant tetrads.  
Notably, these manifolds, while being geodesically incomplete, do not exhibit any singularity either in the curvature two-form components or in the lower-dimensional curvature scalars defined in eq.(\ref{Rbar}). 


For $\sigma=+1$ at $|v|< \epsilon$, on the other hand, there exists no timelike coordinate within the core. This also implies that no timelike or null geodesic can live in this region, and any such curve starting at some distance $v$ with $|v|>\epsilon$ must terminate at the phase boundary $v=\pm \epsilon$. As it is, the spacetime-bridge solution in this case also consists of two causally disconnected `universes'.

\section{Deflection of light in a spacetime-bridge}


Let us assume that a light ray starting from $v\rightarrow \infty$ ($v\rightarrow -\infty$) comes within a (minimum) distance $v=v_0>\epsilon$ ($v=v_0<-\epsilon$) from the centre of the core and then gets deflected away. 
The distance from the origin $v=0$ may be measured in terms of a radial coordinate defined as:
\begin{eqnarray}\label{R}
R=f^2(v)+ 2M~.
\end{eqnarray}
The demand that the origin of this radial coordinate should coincide with $v=0$ fixes one of the parameters among $M$ and $\epsilon$ in terms of the other. This leaves only one free parameter (the size of the degenerate core) which characterizes a spacetime-bridge solution.
In terms of $R$, the null geodesic equations at $|v|>\epsilon$ (at the equatorial plane) read:
\begin{eqnarray}
\dot{R}^2=E^2-\frac{L^2}{R^3}(R-2M),~\dot{\phi}=\frac{L}{R^2}
\end{eqnarray}
Eliminating the affine parameter $\lambda$ from these equations, we obtain the orbit equation as:
\begin{eqnarray}
\left(\frac{dR}{d\phi}\right)^2=\frac{E^2 R^4}{L^2}-R(R-2M)
\end{eqnarray}
In these coordinates, the equations above are of the same form as the photon orbit in the Schwarzschild geometry. Under the assumption that the size of the bridge $\epsilon$ is very small compared to the radial distance $R_0$ at the closest approach, the solution to this equation can be written as:
\begin{eqnarray}\label{Rsoln}
\frac{1}{R}\equiv \frac{1}{f^2(v)+2M}\approx  \frac{1}{R_0}\left[\sin\phi+\frac{M}{2R_0}\cos 2\phi+\frac{3M}{2R_0}\right]
\end{eqnarray}
The closest approach to the bridge corresponds to the turning point $\frac{dR}{d\phi}=0$, which in the small $M$ approximation leads to $R_0\approx \frac{L}{E}$.

 Next, let us align the axes such that at the equatorial plane, the angle $\phi$ is measured anticlockwise from the horizontal axis. The ray is assumed to make an angle $\phi=-\phi_1$ at the initial infinity and $\phi=\pi+\phi_2$ at the final infinity. If the bending (angle) is small, then as a straightforward consequence of eq.(\ref{Rsoln}), we obtain the total angle of deflection to be:
 \begin{eqnarray}
\Delta \phi\equiv \phi_1+\phi_2 \approx \frac{4 M}{R_0}=2\frac{R(|v|=\epsilon)}{R(|v|=v_0)}
\end{eqnarray}
Evidently, from an observational measurement of this angle, the ratio of the radii of the degenerate core and of the distance of closest approach could be determined. 

One may contrast this interpretation with the case of a light ray barely grazing past the surface of a spherical body with a massive core. The corresponding bending angle there encodes the ratio of its mass (Schwarzschild radius) and its radius. It is important to keep this distinction in mind, since notions such as a `mass' or the `radius' of a surface within which the mass is confined do not exist for the spacetime-bridge solutions. 

The result obtained in this section is manifestly independent of the explicit functional form of $f(v)$. The discussion above could be of generic interest in the context of spherically symmetric spacetimes where patches with invertible and noninvertible tetrads are sewn together \cite{kaul,kaul1,sengupta}. 
The important lesson is that from the perspective of an asymptotic observer, the effect on the light ray due to a source with a degenerate (empty) core is indistinguishable from that of one with a massive core. 

\section{concluding remarks}

Here we have  analysed the causal structure and completeness properties of the two-sheeted spacetime-bridge geometries, which are explicit solutions of first order gravity theory.  While these manifolds are geodesically incomplete and are not time-orientable, they are free of any curvature singularity in a precise sense. Each of these solutions consist of two causally disconnected spacetime sheets, which may be interpreted as two `universes' separated by a region with degenerate tetrads.

We have also presented a brief analysis on the deflection of light propagating in a bridge geometry. 
 The results have the same quantitative form as in the case of a Schwarzschild spacetime, albeit with a different interpretation. The angle of deflection in this case depends on the size of the degenerate core, which is the only free parameter in these solutions analogous to the Schwarzschild mass. 
 
It is difficult not to notice another intriguing prospect, where a double-sheeted bridge spacetime  could be thought to supercede a black hole geometry. This would necessarily imply a modification of the standard picture of a singular black hole interior, which would now be replaced by the `nonsingular' bridge.  This may be contrasted with some of the interesting proposals 
invoked in order to resolve the celebrated information loss paradox \cite{hawking}, where the underlying quantum states are expected to conspire in  a way that leads to the disappearance of the horizon.
With the bridge-geometries, however, the possibility of a new structure replacing the horizon and its interior could be realized in a purely classical setting. In fact, these considerations apply equally well to the degenerate extensions of the Schwarzschild exterior, as is implicit in their construction \cite{kaul1}. We anticipate that such solutions to the field equations \cite{sengupta,kaul1} could provide a fertile testing ground for ideas related to the information loss problem or to the physics of gravitational collapse in general.

 \vspace{.1cm}

\acknowledgments 
Discussions with Sayan Kar have been enlightening. The comments of Sukanta Bose, Naresh Dadhich and Romesh Kaul on a preliminary version of the draft are also gratefully acknowledged. I feel indebted to the Theory group, IACS, Kolkata for the warm hospitality during a visit in December 2017 when part of this work was being completed. This work is supported by the grant no. ECR/2016/000027 (ECR award), SERB, Department of Science and Technology, Government of India.

\end{document}